\documentclass{article}
\begin{document}
%  Article info
\title{The Oklo Constraints on Alpha-Decay Half-Lives}
\date{June 30, 2003}
\author{E. Chaffin and J. Molgaard}
\maketitle
\begin{center}
Physics Department, Bob Jones University, Greenville, SC 29614
\end{center}
\begin{abstract}
The Oklo data constrain the depth of the nuclear potential well to a small margin of error, determined by various unknowns, such as the reactor temperature. However, we show that even these small variations could allow the U-238 half-life to vary by more than one order of magnitude.
\end{abstract}
\vspace{0.5in}
\begin{center}
\vspace{0.5in}
\end{center}
\newpage

\section{Introduction} 
Beginning with Shlyakhter [1] in 1976, Oklo natural reactor data have been used by many authors [2-6] to put limits on the time variation of "constants" such as the fine structure constant and the strong coupling constant. As noted by Damour and Dyson [2], such constraints generally refer to the value of the "constants" at the time of the reactor shut down, since most prior evidence is diminished by the reaction process. Also, it cannot be ruled out that two or more factors may change concurrently, and one may cancel the effect of the other on whatever we are able to calculate for comparison with experiment.

In this work, we will use the constraints provided by the samarium-149 cross section to provide a self-consistent limit on the value of the uranium-238 decay constant at the time of reactor shutdown. This implies a limit on possible differences between the strong coupling constant at the time of reactor shutdown and today. Pierronne and Marquez [7,8] developed the complex eigenvalue method for calculating alpha-decay half-lives by matching the wavefunction of the alpha particle, and its derivative, across a boundary between a square well representing the nuclear interior and a coulomb potential representing the exterior of the nucleus outside the range of nuclear forces. Chaffin, Banks, Gothard, and Tuttle [9,10] have further developed this method into {\itshape Mathematica\/} notebooks permitting easy calculation of relevant variations.

Experimental data on alpha particle scattering [11-14] have long been known to validate a cluster model in which an alpha particle moves in the average field provided by the rest of the nucleus. Four-nucleon transfer reactions were observed on target nuclei that are the {\itshape daughters\/} of an alpha emitter. The comparison of reaction rates with the corresponding alpha-decay rates indicates the tendency for the four transferred particles to form a true alpha particle. The data give evidence that nodes in the alpha particle wavefunction do in fact exist. When the nuclear potential seen by the alpha particle is represented by a simple square well, the effect of changing the depth of the well is to change the alpha particle wavefunction. Usually this change is slight, but at certain critical values the number of nodes in the wavefunction can change precipitously, with a corresponding change of more than an order of magnitude in the decay constant. Hence, we find that, although the Oklo data provide Shlyakhter's very stringent limits on the variation in, for instance, the Sm-149 absorption cross-section for neutrons, this may not translate into such stringent limits on nuclear alpha-decay half-lives.

In section 2 of this paper we will present the method of calculation of the Samarium cross-section at the time of the Oklo reactions, and indicate how this may be used to find the depth of the square well representing the nuclear potential. Then in section 3 we review the Pierronne and Marquez [8] method for calculating the decay constant from the depth of the well and other parameters. We then present the results of a calculation for the case of alpha-decay of U-238. 
\vspace{0.5in}
\section{Using Oklo Data to find the Samarium-149 Cross-Section}
We have used actual data of element concentrations from Oklo samples to find the implied Sm-149 absorption cross-section for neutrons. Procedures very similar to those of Damour and Dyson [2] and Fujii {\itshape et al.\/}. [3] were used. As Fujii {\itshape et al.\/} explained in their Appendix B, coupled linear differential equations may be written down for the time derivatives of the Nd-143, Sm-147, U-235, and U-238 concentrations. These equations may be solved for the restitution factor $C_{rz}$, epi-thermal index $r_{rz}$ , and the neutron fluence appropriate for each sample. For each sample the ratio k = ${N_{235}^{0}}$/$N^{nat}$, the ratio of the initial U-235 abundance to the natural abundance of samarium, is also found using Fujii {\itshape et al.\/} 's procedure.

As explained in the Damour and Dyson [2] and/or Fujii {\itshape et al.\/} [3] papers, coupled linear differential equations may also be written for the Sm-147, Sm-148, and Sm-149 concentrations. To obtain a working equation we use equation (16) of Fujii {\itshape et al.\/} [3], corrected for a typographical error in the subscript for the Sm-149 absorption cross-section, yielding the result for the abundance ratios at the end of the reactor activity. If the time for the end of reactor activity is $t_{1}$, and $N_{147}$ , $N_{148}$, and $N_{149}$ are the concentrations of Sm-147, Sm-148, and Sm-149, respectively, the appropriate equation is:
\vspace{0.25in} \large
\begin{equation}
\begin{array}{l}
\frac{{N_{147} (t_1 ) + N_{148} (t_1 )}}{{N_{149} (t_1 )}} = \\ 
\frac{{\frac{{\hat \sigma _{f235} kY_{147} }}{{\hat \sigma _a }}\left[ {1 - \exp \left( { - \hat \sigma _a \hat \phi t_1 } \right)} \right] + \left( {R_{147}^{nat}  + R_{148}^{nat} } \right)}}{{\frac{{k\hat \sigma _{f235} Y_{149} }}{{\hat \sigma _a  - \hat \sigma _{149} }}\left[ {\exp \left( { - \hat \sigma _{149} \hat \phi t_1 } \right) - \exp ( - \hat \sigma _a \hat \phi t_1 )} \right] + R_{149}^{nat} \exp ( - \hat \sigma _{149} \hat \phi t_1 )}} \\ 
\end{array}
\vspace{0.25in}
\end{equation}
\normalsize
Here the R's are the relative fractional natural abundances of the isotopes of samarium indicated by the subscripts, $Y_{147}$ and $Y_{149}$ are the fission yields of Sm-147 and Sm-149, respectively. Using this equation, we used an iteration procedure called FindRoot in {\itshape Mathematica\/} to obtain the samarium-149 absorption cross-section.

Results are given in Table I, where they are compared to Fujii {\itshape et al.\/} 's results.

The table gives the cross-sections obtained for samples inside the core of reactor zone 10, and average cross-section of $90.29 \pm 7.41$ kilobarns. The errors are the one standard deviation statistical errors. The temperature of the Oklo ores at the time of the reactions is unknown but thought to have been around 300 to 400 degrees Celsius. This temperature would, of course have varied over the lifetime of the nuclear reactor, and is an important systematic source of error in determining, from the data, the Sm-149 cross-section at the time of the reactions.
\section{Finding the U-238 Decay Constant}
In a simple nuclear model in which a square well is used for the potential seen by the alpha particle inside the nucleus, the Sm-149 cross-section is related to the depth of the potential well. The cross-section as a function of energy is given by the Breit-Wigner shape:
\begin{equation}
\sigma _r  = A\frac{{\Gamma _n \Gamma _\gamma  }}{{\left( {E - E_r } \right)^2  + \Gamma _{tot}^2 /4}}
\end{equation}
In this expression the widths $\Gamma$ do not depend directly on the depth $V_{0}$ of the potential well, but the resonant energy $E_{r}$ is a constant minus $V_{0}$, as was shown, for example, by Weber, Hammer, and Zidell [15]. The spread in values of the samarium cross-section found in section 2, thus can be related to a spread in values of $V_{0}$. For U-238, the half-life of $4.47 \times 10^9$ years is reproduced by a $V_{0}$ of 110.5 Mev. From the data of section 2, the integrated Sm-149 cross section is found to be $90.3 \pm 7.4 $ kilobarns (see Table I). The samarium resonant energy is ${E_r}$ =0.0958325 eV and the corresponding one standard deviation spread is $\pm 0.0102$ eV. While this spread might seem small, the small change in $V_{0}$ can lead to a large change in the alpha-decay half life.

To test the variability of the decay constant, {\itshape Mathematica\/} notebooks were written. For the square well potential on the inside of the nucleus, the {\itshape Mathematica\/} notebook gave answers essentially equivalent to those of Chaffin, Gothard and Tuttle [10], which used a harmonic oscillator potential for the interior region, where the nuclear potential is felt by the alpha particle. In the course of this work, it was discovered that, as the nuclear potential well depth is changed, and the nuclear radius changes slightly, it is possible to have a sudden change in the number of nodes of the real part of the alpha particle's wavefunction. This was modeled for both the harmonic oscillator and square well potentials, with nearly the same results for either notebook.

The change in the number of nodes causes the probability of tunneling to change by about a factor of 13.5, as shown by the discontinuity in the graph of Figure 2. The graph shows a large, precipitous rise in the value of the decay constant $\lambda$, at certain critical values of $V_{0}$, similar to the postage-stamp function.

\section{Conclusions}
The calculations of the decay constant versus well depth $V_{0}$ show that even a small changes in the strong coupling constant could result in a U-238 decay constant change by a factor of 13.5, more than one order of magnitude. The effect is tied to the change in the number of nodes in the alpha-particle wavefunction when the well depth passes critical values.
\\
\textbf{References}

[1] Shlyakhter, A. I., Nature 264, 340 (1976).

[2] Damour, T. and F. J. Dyson, Nuclear Physics B480, 37-54 (1996).

[3] Fujii, Y., A. Iwamoto, T. Fukahori, T. Ohnuki, M. Nakagawa, H. Hidaka, Y. Oura, and P. M\"{o}ller, Nuclear Physics B 573, 377-401 (2000).

[4] Irvine, J. M., Contemporary Physics 24(5): 427-437 (1983).

[5] Irvine, J. M., Philosophical Transaction of the Royal Society A310, 239-243 (1983).

[6] Irvine, J.M. and R. Humphreys, Progress in Particle and Nuclear Physics 17, 59-86 (1986).

[7] Marquez, L. Phys. Rev. C25(2), 1065-1067. (1982).

[8] Pierronne, M., and L. Marquez, Zeitschrift f\"{u}r Physik A286, 19-25 (1978).

[9] Chaffin, E. F., and D. S. Banks, nucl-th/0206020.

[10] Chaffin, E. F., N. W. Gothard, and J. P. Tuttle, nucl-th/0105070.

[11] Anyas-Weiss, N., J.C. Cornell, P.S. Fisher, P. N. Hudson, A. Menchaca-Rocha, D.J. Millener, A.D. Panagiotou, D.K. Scott, and D. Strottman, Physics Reports C12(3), 201-272. (1974).

[12] Buck, B. Nuclear-structure information from several-nucleon transfer reactions, in {\itshape Nuclear spectroscopy and nuclear reactions with heavy ions\/}, International School of Physics Enrico Fermi course 62. (North Holland, Amsterdam, 1976).

[13] Davies, W. G., R.M. DeVries, G.C. Ball, J.S. Forster, W. McLatchie, D. Shapira, D., J. Toke, and R.E. Warner, Nuclear Physics A 269, 477-492 (1976).

[14] Wildermuth, K., and W. McClure, {\itshape Cluster representations of nuclei \/} (Springer tracts in modern physics No. 41). (New York: Springer-Verlag, New York, 1966).

[15] Weber, T.A., C.L. Hammer, and V.S. Zidell, American Journal of Physics 50(9): 839-845 (1982).

[16] Green, A. E. S. and K. Lee, Phys. Rev. 99, 772-777 (1955).
\newpage
\textbf{Figure Captions} 

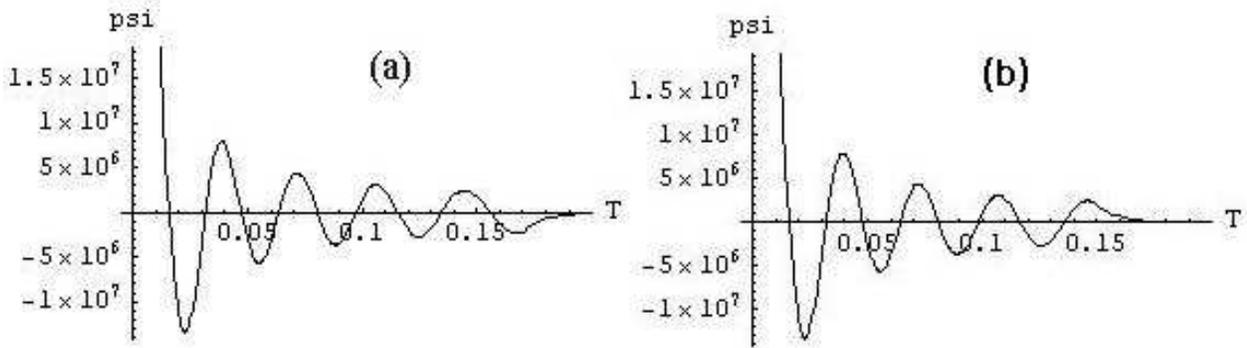
Figure 1. Sudden change in the number of nodes. The harmonic oscillator wave function for well depths of 58 MeV (a) and 54 MeV (b). The x-axis is the radial coordinate of the alpha particle, $T= \rho /(2 \eta )$, where $\rho$ and $\eta$ are defined in Green and Lee [16]. Figure 1a shows the harmonic oscillator wave function for a well depth of 58 MeV. Figure 1b shows what happens when the well depth is changed to 54 MeV, without changing the alpha particle energy. If one counts the number of nodes in Figure 1a, there are nine, not counting the ones at zero and infinity. For Figure 1b, there are eight nodes, a reduction by one.
\par
\vspace{0.5in}
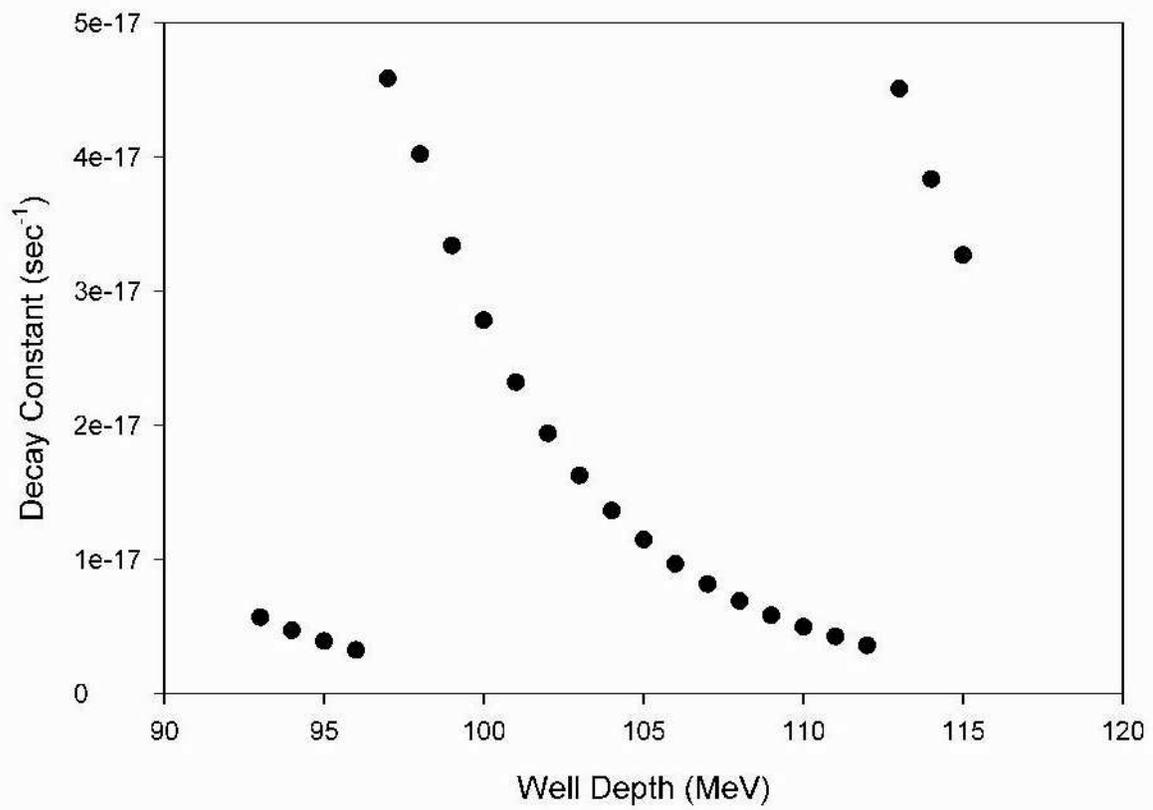
Figure 2: The Decay Constant $\lambda$ of U-238 versus Well Depth.
\newpage
\begin{table}
\begin{center}
\begin{tabular}{||l|r|r||}
\hline
Sample &$\hat\sigma_{149}$ (kb) &Fujii {\itshape et al.\/} results \\ \hline
SF84-1492 &98.1 &99.0 \\ \hline
SF84-1485 &83.0 &83.8 \\ \hline
SF84-1480 &95.0 &96.5 \\ \hline
SF84-1469 &85.0 &85.6 \\ \hline
standard deviation &7.41 &7.64 \\ \hline
average &90.3 &91.2 \\ \hline
\end{tabular}
\caption{Sm-149 Effective Cross Sections}\label{Ta:first}
\end{center}
\end{table}
\end{document}